\documentstyle[prb,aps,preprint,epsfig]{revtex}
\tightenlines
\begin{document}

\title{A set-up for measurement of low frequency conductance fluctuation (noise) using digital signal processing techniques}
\author{Arindam Ghosh,$^{a}$ Swastik Kar,$^{b}$ Aveek Bid and A. K. Raychaudhuri$^{c}$}

\address{Department of Physics, Indian Institute of Science, Bangalore 560 012, India}
\date{\today}
\maketitle

%\twocolumn[\hsize\textwidth\columnwidth\hsize\csname
%@twocolumnfalse\endcsname

\begin{abstract}
We describe a set up for measurements of low frequency (1 mHz $\lesssim f \lesssim 20$ Hz) conductance fluctuations in a  solid conductor.  The set-up uses a five probe a.c. measurement technique and extensive digital signal processing to reach a noise floor down to $S_{v}(f) \leq 10^{-20}$ V$^{2}$Hz$^{-1}$. The set up also allows measurement of noise using an a.c. in presence of a superimposed direct  current. This feature is  desirable in studies of electromigration damage or in systems that show non-linear conductivity. In addition, we describe a scheme which allows us to obtain the probability density function of the conductance  fluctuation after subtracting the extraneous noise contributions (background) from the observed noise. The set-up has been used for conductance fluctuation measurement  in the temperature range 1.5 K $\lesssim T \lesssim 500$ K in the presence of magnetic fields. We present some representative data obtained by this system.
\end{abstract}

\noindent{\small $^a$Present address: Cavendish Laboratory, University of Cambridge, Madingley Road, Cambridge, UK.}\\
{\small $^b$Present address: Physikalisches Institute, Universit\"{a}t Karlsruhe, 76128 Karlsruhe, Germany. }\\
{\small $^c$email: arup@physics.iisc.ernet.in}

\newpage
\noindent

\section{Introduction} 
Study of conductance fluctuation is a valuable tool to study low frequency dynamics in condensed matter systems~\cite{vdz,DH,weissman1,weissman2,bkj}. The conductance fluctuations  in a current carrying conductor often show a power spectrum $S_V(f)$ $\propto 1/f^{\alpha},\alpha \approx 0.8-1.2$ and this is referred to as ``$1/f$ noise". A proper application of $1/f$ noise  spectroscopy  can be useful in different areas of materials science ~\cite{akr1}. Measurement of conductance or resistance fluctuations often  involve detection of voltage changes of the order of 0.5nV-5 nV across the sample. The sample (the resistor) is usually current biased by a low noise source and the voltage fluctuation($\delta v(t) =
v_{0}(t)-\langle v\rangle$, where $\langle v\rangle$ is the average voltage) is the manifestation of conductance
fluctuation. $\delta v(t)$ is the physical quantity that is measured and the measurement over a time period makes a time series. This is depicted schematically in figure~1. 
The most common though not the only way to represent the noise is through its spectral weight $S_{v}(f)$ defined as:

\begin{equation} S_{v}(f) =
\lim_{T\rightarrow\infty}(1/2T)[\int_{-T}^{T} dt \delta
v(t)\exp(-2i\pi ft)]^{2} \label{defination1}
\end{equation}

Power spectrum is also the Fourier transform of the auto-correlation function $c(\tau)$ defined as:

\begin{equation}
C(\tau) = \lim_{T\rightarrow\infty}(1/2T)\int_{-T}^{T} dt \delta
v(t+\tau) \delta v(t) \label{defination2}
\end{equation}

The core problem of experimental noise spectroscopy is the elimination  of the contribution  of noise from sources other than the sample using both software and hardware techniques.  The unambiguous and reproducible determination of the spectral power of the conductance fluctuation  from the sample is a particularly demanding experiment, more so  when the sample has low noise levels, since the measurement instrument noise itself has a $1/f$ character.

The conductance fluctuations (or noise) measurements have two distinct aspects. The first relates to the hardware, involving elimination of the background noise  arising from the environment, temperature control and other related issues. The extent of temperature control determines the low frequency range of reproducible data particularly for $f \leq$ 100 mHz. The reduction of external noise coming mainly from electromagnetic interferences, ground loops and impedance mismatch requires adoption of a few special techniques. Background noise also involves such extraneous sources such as the contact noise which need be eliminated or minimized.

The second and equally important aspect of noise measurement is the software aspect that involves various aspects of the data reduction involving digitization, digital filtering, fast fourier transforms (FFT)and related techniques. The data reduction technique  involves use of  digital signal processing (DSP). Use of modern DSP techniques along with the associated hardware leads to significant rejection of external noise.  We report in this paper how such techniques can be effectively implemented in practice in the context of noise measurements. The realization of the DSP techniques described in  this paper allows a spectral noise floor of lower than  $S_{v}(f) \leq 10^{-20} V^{2}Hz^{-1}$ and allows measurements down to 1 mHz. The computational speed and power available in modern PC's can be effectively utilized for this purpose.
\begin{figure}[h]
\begin{center}
\epsfig{figure=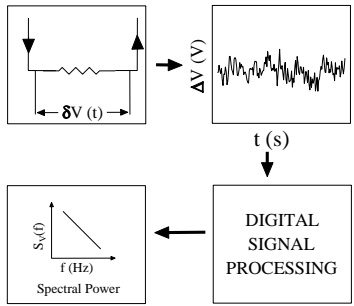,width=6cm,height=6cm,clip=}
\caption{Schematic representation of the basic technique to obtain power spectra from the time series. The voltage fluctuation $\delta v(t) = (v_{0}(t)-\langle v\rangle$), across the current biased resistor is measured.This is the 4-probe configuration.}
\end{center}
\end{figure}

Our set-up is based on the five-terminal a.c. noise measurement technique described below. We have also improvised the technique so that one can measure the noise by a 5 probe a.c technique in the presence of a superimposed d.c. This is useful for studying noise during electromigration stressing or in systems with non-linear conduction.

While the estimation of power spectrum is the most common method to quantify fluctuations, it is not the only way. A useful way to present the data is through the probability density function (PDF) of the fluctuations, a statistical distribution of data estimated from real time resistance fluctuations. Often the real time signal is corrupted by extraneous signal that is difficult to isolate from the fluctuation to be studied. We also describe a scheme by which the extraneous signal can be isolated and one can obtain the PDF of the fluctuation from the sample. Direct measurement of the PDF is a way to find out whether the fluctuation being studied is Gaussian. Another way to find the deviation from Gaussian behavior is through the measurement of higher order correlation functions which are expressed through ``second spectra" (defined later on). We also describe  how the data acquired can be used to find the second spectra.

We have used this apparatus in a number  of experiments over the years~\cite{ag1,ag2,ag3,ag4,ag5,ag6,sk1,akr2,ag7,sk2,ab1,sk3}. This report describes in  details the measurement methods used in the publications stated above. We also provide a selected representation of the systems studied using this technique.Though the main focus of the paper is on applications of DSP methods to noise measurements, we include a brief description of the hardware as well.

\section{The perspective}

In this subsection we describe briefly the most common methods of noise measurements for the sake of completeness. This is to put the present investigation in a proper perspective. As mentioned before, the quantity that one experimentally observes is the fluctuating physical quantity (say a voltage $v(t)$) as a function of time around a mean defined as $\delta v(t) = v_{0}(t)-\langle v\rangle$, where $\langle v\rangle$ is the mean voltage. The main information is contained in this time series $\delta v(t)$.The most popular way to represent the noise is through the power spectrum $S_V(f)$ of
the measured fluctuation $\delta v(t)$ defined above in equation~1. The frequency range for which the spectral power is to be determined depends on the time scale of the fluctuations being investigated which in turn depends on the dynamics associated with the underlying physical process that causes the fluctuation. In general, if the dynamics is occurring at a time scale $\tau$, it will affect the power spectrum around the frequency range $f\sim(1/2\pi \tau)$. The relaxation process that can cause the conductance to change with the characteristic time $\tau$ makes a contribution  to the power spectrum $S_{v}(f)$ of the voltage fluctuation which  is given by a Lorentzian  $2\tau /(1+(2\pi f\tau)^2$. For processes with a distribution of relaxation time $F(\tau)$ the power spectrum is a superposition of Lorentzians given as~\cite{DH}:

\begin{equation}
S_V(f)=\int_{0}^{\infty} d\tau F(\tau) 2\tau/(1+(2\pi f\tau)^2
\label{defination3}
\end{equation}
If there is a smeared kinetics then with a proper (and physically realizable)choice of $F(\tau)$ it possible to obtain a power spectra $S_V(f) \propto 1/f^{\alpha}$ ($\alpha \sim 1$).
In the experimental measurement of $1/f$ noise it is  important to ensure that the noise originates from the sample since many of the associated measuring electronics  have noises that also have a ``$1/f$" power spectrum.

A common way to measure $1/f$ noise is to bias the sample with a constant current. In general, a 4-probe configuration is used where the biasing current is applied through two separate leads just like in 4-probe resistivity measurement (depicted in figure~1). The biasing current being d.c , the steady voltage drop across the sample is blocked off by a capacitor bank forming a high pass filter. The fluctuating voltage across the materials to be studied is then amplified by a low-noise preamplifier (for samples with relatively higher resistance) or a low noise transformer for more metallic samples. The power spectrum can be obtained using a commercial spectrum analyzer or the time series, $\delta v(t)$, can be digitized and the power spectrum be obtained from the FFT of the auto-correlation function of the time series. The ease of use  has made the spectrum analyzer the most commonly used  method although it may not always be the best option.  In order to separate out the contribution of the background to the measured the $S_V(f)$,  the $S_V(f)$ is also measured with zero bias current and then subtracted out from the total measured  $S_V(f)$ . The difference gives the noise due to the sample under measurement.

The 4-probe d.c method suffers from two very important drawbacks: (a) the 1/f noise of the preamplifier and the associated electronics adds to the background and raises the noise floor of
the measurements and (b) the background estimation can be very faulty because it can shift during the measurement itself since long data collection times are necessary for measuring the power
spectra at lower frequency. Generally, if the power spectrum $S_V(f)
\geq 10^{-17} V^{2}Hz^{-1}$ and the background is relatively small and stable
then the d.c method may be used.

Most of  the drawbacks of the d.c technique is overcome by use of a.c technique. An important realization of the a.c technique has been originally proposed by Scofield~\cite{scoff1}. In this technique, the sample is biased by an a.c carrier and the signal is demodulated by a lock-in amplifier. This shifts the detection frequency to a region where the preamplifier has very low $1/f$ component. Typical carrier frequencies are around 1 KHz. The choice of the carrier frequency is to operate on  region of low noise figure of the preamplifier and the lock-in amplifier. The required information on the noise contour is generally given in the manuals of lock-in amplifiers and preamplifiers. The out put signal from the lock-in amplifier is then digitized. In this case the bandwidth of the output low-pass filter of the lock-in amplifier will decide the upper-band limit of the spectrum. Generally the carrier frequency is much larger than the  bandwidth of the power spectrum measured. For measuring power spectra with $f \leq 10^{2}$ Hz this is a convenient technique. In addition, if the signal is recorded at two phases, it can be shown that the in-phase part will give a power spectrum that is a sum of both
background and ``signal" while the out-of-phase component gives the background only.~\cite{scoff1} The power spectrum of the noise from the sample (which we refer to as ``signal" in the context of our work) can be obtained by subtracting the background, which is measured simultaneously. The simultaneous measurement of the background is the greatest advantage of the a.c technique~\cite{scoff1}. A closely linked method has been described by Verbruggen et.al~\cite{verbruggen}. The main source of irreproducibility in quantitative measurement of $S_V(f)$ is the improper estimation of the background contribution to the spectral power and its subtraction from the total measured power. It is not only important to reduce the back ground spectral power but also to have a good quantitative measure of it. By simultaneous measurement of back ground and the total power, the  a.c measurement eliminates a good part of the problem. For the sake of completeness we briefly mention the mathematical basis of the noise measurement procedure by the a.c method using a lock-in-amplifier ~\cite{scoff1}.  The power spectral density of voltage fluctuations at the output of the lock-in-amplifier is given by,

\begin{equation}
\label{liaop}
\Sigma_V(f,\delta) \simeq G_0^2\,[S_V^0(f_0-f) +
   (I_{rms}^2) S_r(f) cos^2(\delta)]
\end{equation}

\noindent where $S_V^0(f)$ is the voltage fluctuation due
background noise, uncorrelated to the samples noise and $S_r(f)$
is spectral density of resistance fluctuations of the sample.
$G_0$ is the product of the gain of the pre-amplifier and that of
the lock-in-amplifier, $f_0$ is excitation frequency,$f$ is the
measurement frequency ($f \ll f_0$), $I_{rms}$ is the rms value of
the biasing current and $\delta$ is phase angle of detection of
the input voltage to the lock-in-amplifier with respect to the
bridge current. The relation above is valid for $f < f_0/2$. If
the pre-amplifier noise and other extraneous noise sources are low
or properly eliminated,the main contribution to the  low frequency
background noise, is mainly the Johnson noise with $S_V^0$$\approx
4k_BTR$~\cite{reif}. The spectral density of voltage fluctuations from the
sample, $S_V(f)$ is related to the spectral density of resistance
fluctuations as $S_V(f) = I_{rms}^2 S_r(f)$.Hence, when $\delta$ =
0, $\Sigma_v(f,0)/G_0^2 = S_V^0 + S_V(f)$, i.e., when the
lock-in-amplifier demodulates the input signal at the same phase
as the bridge current, the output contains noise power from both
the sample and that of the background . Secondly,when $\delta$ =
$\pi$/2, $\Sigma_V(f,\pi/2)/G_0^2 = S_V^0$, i.e., detection at
90$^\circ$ gives only the background noise in the output.

An improved version of the a.c. method is the 5-probe
technique, where the central electrode is grounded ~\cite{scoff1,ag8}. This is shown as
schematic in figure~2. 
\begin{figure}[h]
\begin{center}
\epsfig{figure=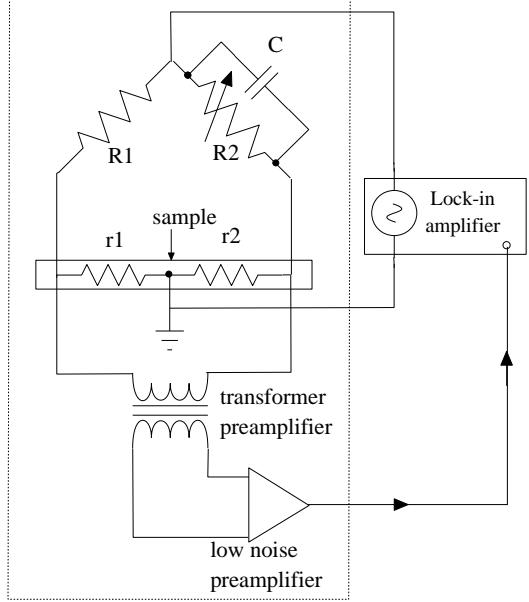,width=8cm,height=8cm,clip=}
\vspace{1cm}
\caption{Schematic representation of the basic technique to obtain power spectra from the time series. The voltage fluctuation $\delta v(t) = (v_{0}(t)-\langle v\rangle$), across the current biased resistor is measured.This is the 4-probe configuration.}
\end{center}
\end{figure}
Grounding the centre effectively divides the sample into nearly two equal arms which form the lower arms of an a.c bridge. Low noise wire-wound resistors form the two upper arms. The upper arm resistances are chosen to be much larger than that of the lower arm (sample) resistance, so that the current through the sample is determined by the upper arm resistors. The bridge can be balanced both in amplitude and phase by shunting one or both the upper arm resistors with small, low noise variable capacitors. In the balanced condition, the signal to the amplifier contains only the information of the fluctuations and not the mean bias across the sample. Thus the signal reaching the amplifier is rather small in amplitude, allowsing one to use large amplification. Thus balancing the bridge can raise the sensitivity many folds. The 5 -probe configuration also reduces the  contributions to noise by temperature drift. This happens because two similar half of the sample forms two arms of the bridge and the drift in resistance due to temperature drift is balanced out. Balancing out the temperature drift contribution allows spectral noise determination to lower frequency ($f \leq$ 10mHz) By working with a carrier frequency $f_0$ at which
the amplifiers have a very low noise  one can measure the spectral
power to  very low frequencies ($\sim$ 1 mHz).  We note that measurement of
$S_V(f)$ to a frequency of 1mHz by  the conventional d.c. methods is
particularly difficult because the measured noise spectral density
is masked by the amplifier noise.

The exact methods to be used for noise measurements depend on the noise level of the measured sample. When the spectral power of the sample noise $S_V(f) \geq 10^{-16}-10^{-17}$ V$^{2}$Hz$^{-1}$, the background is usually not a problem and simple d.c method will suffice. However, noise measurements in relatively clean systems can involve measurement of spectral power as low as $10^{-20}$ V$^2$Hz$^{-1}$ or even lower. In such situations, the contribution of sample ``conductance fluctuations" in to the total spectrum maybe visible only at very low frequencies, the region of the spectrum at higher frequencies being dominated by background noise (see, for example, figure 14(b)). To botain the complete spectrum of conductance fluctuations, the background level has to be carefully estimated and subtracted. For low resistance samples, where a transformer premaplifier can be used, a good way to ensure that the background noise is minimized is to compare its power spectra  with the expected power spectra of the thermal noise ($4k_BTR$). If the power spectra of the background is comparable to $4k_BTR$ (or within a reasoanable range of it), one can conclude that a good elimination/minimization of the background noise, along with a satisfactory level of temperature control, has been achieved.

In this report, we describe how we have achieved cleaner and more reproducible noise measurement using DSP techniques. In particular, in this paper we describe how our set-up attends to the issue of digitizing techniques and processing of the data after digitization.

\section{The setup}

In this subsection we briefly describe the hardware aspects of our set-up. As an example of a simple low-noise   practical realization of the Wheatstone bridge for 5-probe set up, we have
used  three ten turn wire-wound pots ~\cite{Bourns} of resistances 20 k$\Omega$, 1 k$\Omega$ and 200 $\Omega$  connected in series as the balancing arm. Alternatively a  decade box  can be used. The other arm consists of a few 5 W, wire-wound,non-inductive resistors. Typically, the arm-resistance was kept $\geq$ 30 times the sample-resistance in most of the experiments. The error voltage across the sample is amplified by a low noise amplifier which can be a shielded Triad (G10)transformer~\cite{triad} or a transformer preamplifier (SR 554)~\cite{srs} for low resistance sample ($<100$ $\Omega$). For sample resistances $>100$ $\Omega$ we used a low noise preamplifier (SR 560)~\cite{srs} . The carrier frequency was chosen in each case to lie in the ``eye'' of the Noise Figure (NF) of the transformer and preamplifier to minimize the contribution of the transformer/amplifier noise to the background noise. (The  eye of the NF of the transformer, preamplifier and the lock-in amplifier are all very similar and lie in the range ~200-1000 Hz. Thus the carrier frequency is chosen in this range). The fluctuating output of the bridge was amplified typically by a factor of 100 and the amplified voltage goes to the input of the lock-in-amplifier. Typical balance attained was around 1 part in $10^4$. We used a SR 830 dual phase DSP lock-in amplifier in our work ~\cite{srs}. The lock-in amplifier (depending on its sensitivity and whether the output expand function of the lock-in is used) can be provide a gain of $10^7$. So the total gain one obtains is in the range of $10^9$.We  note that the use of maximum possible gain depends on how well the bridge is balanced and that the lock-in amplifier is not overloaded.

The amplifier noise at low frequencies ($< 10$ Hz)is rather high,decreasing as the frequency increases. It reaches a minimum  at $\sim$ 200 Hz - 900 Hz and at an input resistance of $\sim$ 1
M$\Omega$ (the NF of  the``eye'' of the amplifier). By choosing a frequency of the biasing current($f_0$) within the above range and stepping up the input impedance of the amplifier by a transformer, the
amplifier contribution to the background noise could be made negligible. We were able to carry out noise measurements to a frequency range as low as 1 mHz using this set-up.

In this context we note the superiority of the  digital
lock-in-amplifiers compared to its  analog counterpart. A low
noise phase sensitive detection is the primary feature of a
digital lock-in-amplifier. The lock in amplifier uses anti-alias
filter at the input stage and all the operations of the phase
sensitive detection, including phase shifting, mixing and low-pass
filtering are done by the on-board digital signal processor (DSP)
at 20 bit accuracy. The digitized input voltage is multiplied with
the digitally computed reference sine wave with extremely low
harmonic content ($\leq -120$ dB). Apart from this, its phase
noise ($< 0.0001^\circ$ at 1 kHz), orthogonality (90$^\circ \pm$
0.001$^\circ$) or harmonic rejection ($< -80$ dB) are important
parameters that help to achieve low noise floor.

We carried out most of our experiments in liquid He cryostats that
necessitate use of long cables. These cables are often source of
large background. We used low noise miniature coaxial cables ~\cite{cooner} for all leads inside the cryostat.Special attention was given to the electrical grounding
of the setup and the shields of the coaxial cables.``Single point
grounding system'' with series connection  was implemented which
gave excellent results at low frequencies. The single ground point
was made on the body of the shielding enclosure which was
connected to the main ground of the building.

Figure~2 shows schematic of the setup enclosed by a double-walled
Faraday box. The walls of the box consists of adjacent sheets of
aluminium and mild steel, both having a thickness of 1.5 mm. The
field amplitude inside the box is reduced by both absorption and
reflection at the shield. The shielding, particularly above 50 Hz,
is primarily achieved by the mild steel sheet due to its high
permeability. At 50 Hz, the skin-depths for mild steel and
aluminium are 0.5 mm and 11 mm respectively, which would lead to a
total attenuation of $\sim 30 - 35$ dB across the 1.5 mm thick
sheet. Experiments done to check this revealed about 2 orders of
magnitude suppression of the power line (and its harmonics)
interference inside the enclosure, which was better than the
expected attenuation. Hence use of 1.5 mm thick sheets as shields
appears to be satisfactory. (Use of the shielded enclosure reduces the contribution of external contributions like the power line from about   $10^{-10}-10^{-11}$
V$^2$Hz$^{-1}$ to $\sim10^{-12}-10^{-13}$ V$^2$Hz$^{-1}$.)

\section{Measurement of noise with a.c in presence of a d.c bias }

It is often required to measure noise in a system in presence of d.c stressing current, e.g. noise measurement during an
electromigration stressing of a sample. Alternately, it maybe of interest to measure noise at different d.c bias values, a particularly desirable feature for systems showing nonlinear conduction. This is because the observed voltage fluctuations
needs to be scaled by the measuring bias and not the d.c. bias, in order to  obtain the correct relative fluctuation $S_{V}/V^2$. If the bias for non-linear conductivity and noise measurements are the same, $S_{V}/V^2$ becomes a rather ill-defined quantity.

We have used a setup in which noise can be measured with a fixed amplitude a.c signal as described above but one can independently apply a d.c bias which does not interfere with the noise measurements. Thus, in our set-up (described below) one can obtain non-linear bias-dependent conductivity, and yet retain a well-defined meaning for  $S_{V}/V^2$. The d.c bias  ($J_{dc}$)  has been applied using a circuit shown in figure~3. 
\begin{figure}[h]
\begin{center}
\epsfig{figure=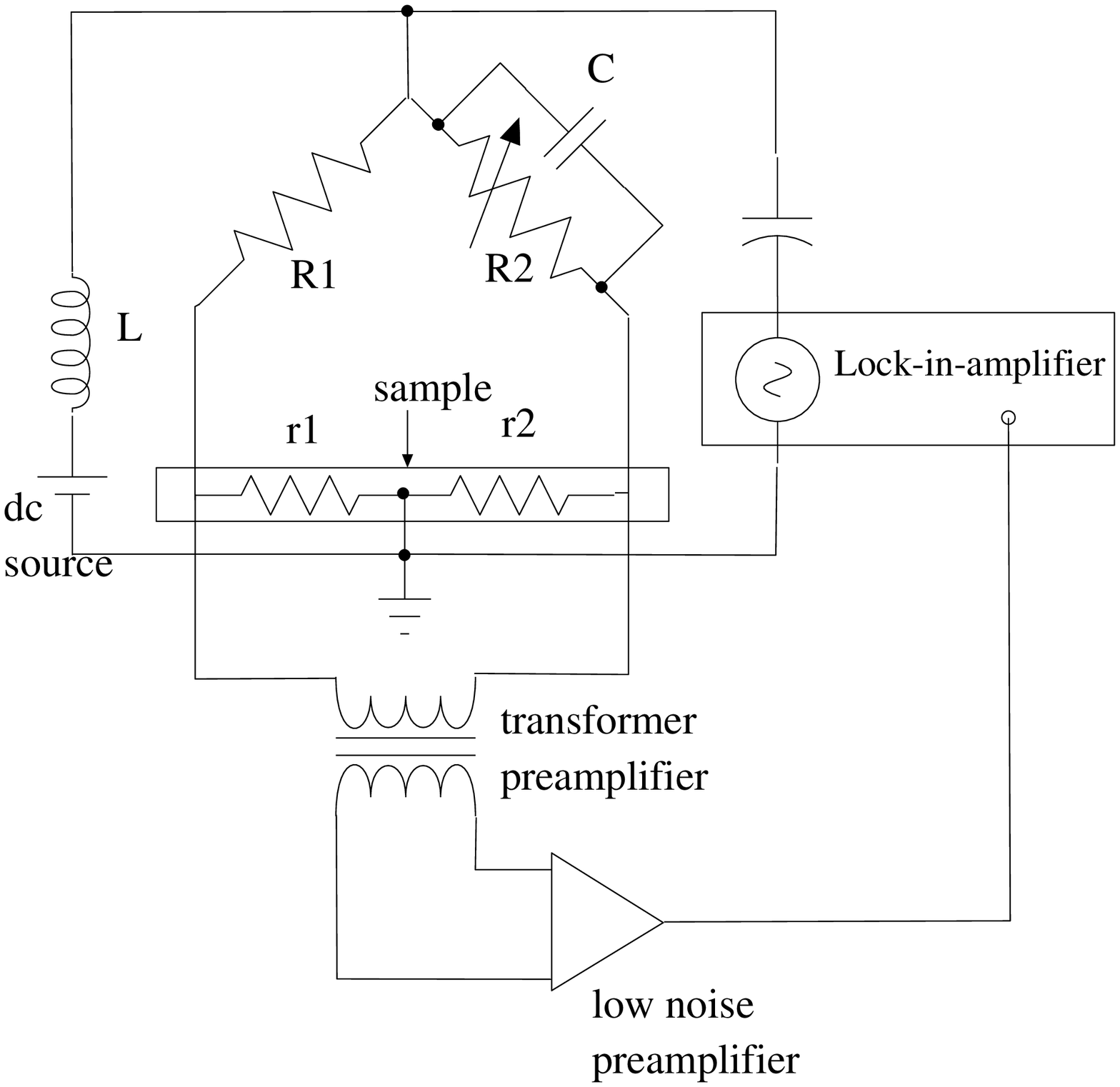,width=6cm,height=6cm,clip=}
\epsfig{figure=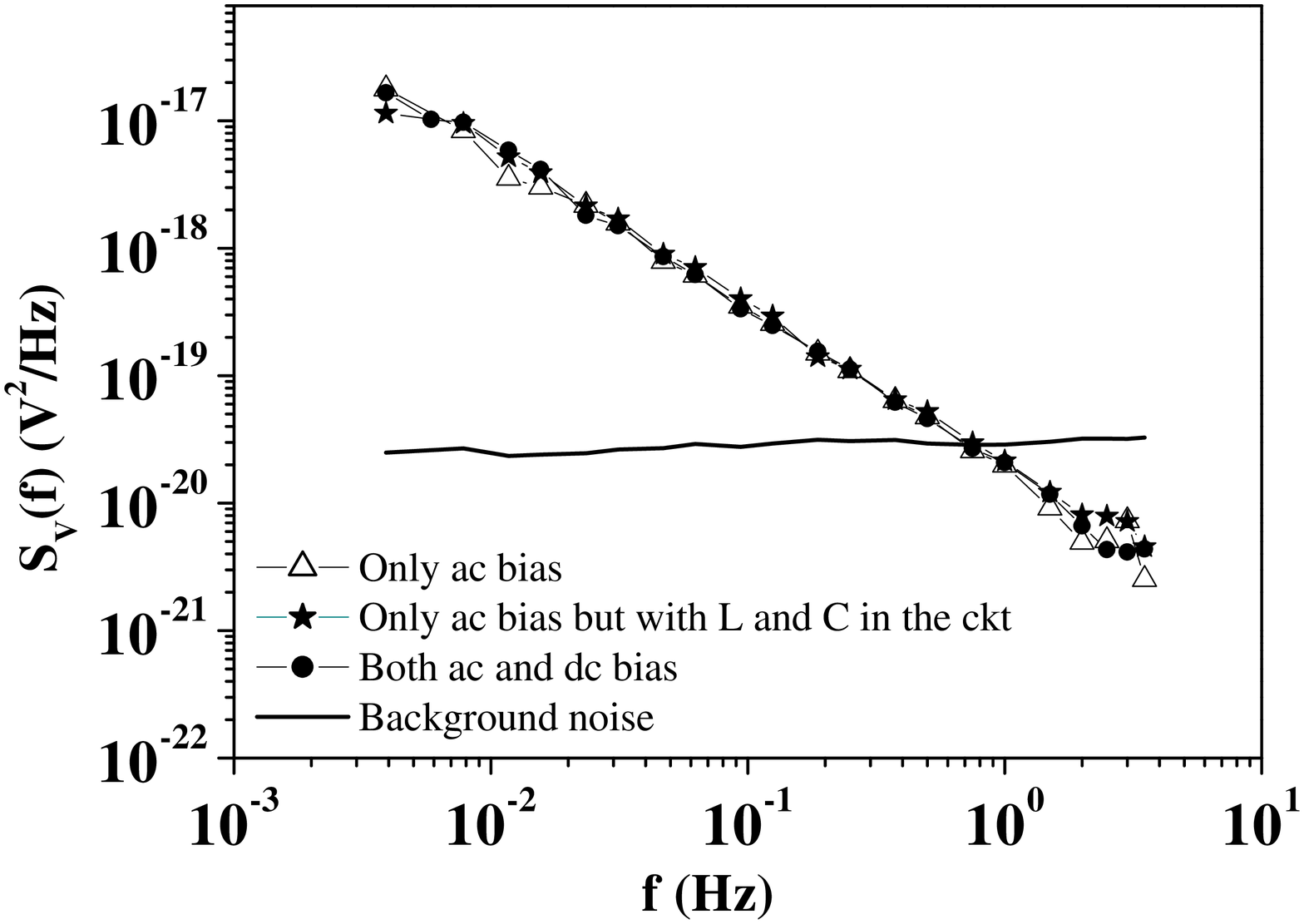,width=8.6cm,height=6cm,clip=}
\vspace{1cm}
\caption{(a) Schematic of the connections when the noise is measured with a.c in presence of a superimposed d.c. Typical air core inductor has $L= 0.8H $ and the silver mica capacitor has $C = 10 \mu F$. (b) Noise data taken with a.c bias with and without superimposed  d.c bias.}
\end{center}
\end{figure}

The capacitors and inductors shown in the figure decouple the a.c and d.c. source and prevent d.c being applied  to the pre-amplifier circuit. It is important to check that the capacitors and inductor does not
affect the gain and the phase of the amplifier and also does not introduce additional features in the power spectra. This was
tested by taking data on a metallic film at room temperature by biasing the circuit using only an a.c signal ($J_{dc}$=0) and taking data with the decoupling capacitors and inductors both present and absent. Data were also taken using a.c and d.c  applied together. In all these cases power spectra obtained were seen to be identical as shown in figure~3.  ( Note : The metallic film has a linear I-V curve and is ohmic in nature.  $J_{dc}$ was kept below the electromigration threshold. )

\section{Digitization and Signal Processing}

A very important aspect of power spectrum estimation without artifacts is the digitization of the signal appearing at the output of the lock-in amplifier, and subsequent digital signal processing to obtain the power spectrum. Two different mechanisms of transferring the voltage fluctuation data to an external storage media (like the computer memory or hard disk) can be  adopted.

In the first procedure, the lock-in-amplifier DSP was triggered by the computer through IEEE 488 interface ~\cite{national instruments} for data acquisition and two 64 Kbyte  in-built buffers of SR830 were used
for temporary storage. However, due to restriction in the
available memory in SR830, a maximum of only 16383 data points
could  be acquired in a single run. On completion, the stored data
was transferred to the computer memory through the IEEE 488 interface card. This method is useful for multiple number of short data-sets in systems which have inherent drifts (e.g. a stressed electromigration film) which tend to overload the lock-in during data storage. The time delay of the data transfer can be used to re-balance the bridge before the next set is stored in the buffer. However, this is also a limitation when the
power spectra is required to be estimated down to very low frequencies, since the averaging for larger timescale datapoints become restricted to only a very few points.

The second method used a 16 bit analog-to-digital converter card (PCL816)~\cite{plc}
which was attached to the computer mother board. The input signals
were digitized  at a definite rate determined by a programable
on-board pacer clock (8254) and directly transfer the data to the
computer memory. Since the available memory space was much larger,
this method was generally used to collect large number of points
($\sim 10^6$ or more) in a single run. No spectrum analyzer was used in
any of the experiments because the computer provides a direct
access to the time series.

As mentioned earlier, noise measurements invariably involve high
degree of interference from environmental sources such as power
line frequencies and its harmonics. Even with extensive grounding and
shielding, only a part of this undesirable interference can be
suppressed. However, for noise measurement purposes, specially
when the measurable noise is $\sim 10^{-20}$ V$^2$/Hz or less, one has to implement a few special DSP techniques to suppress any contribution at power line frequency or its harmonics. The rejection of excess interferences depend on several parameters, as described below.

Typically, the intended sampling rate was restricted  to $\Delta \leq 32$ Hz. This determined the required time constant
($\tau_{lock}$) of the filter at the output of the lock-in-amplifier as, $\tau_{lock} \approx 1/(2\pi\Delta)$. The
roll-off of the filter was kept at a maximum available value of 24 dB/oct. An important issue is the choice of the value of
$\tau_{lock}$. The problem involved may be appreciated as follows. For instance, digitizing the data at 32 Hz with $\tau_{lock}
\approx$ 3 msec, i.e., a pass-band width  of $\approx$ 53 Hz ($=1/2\pi\tau_{lock}$), results in high degree of aliasing. 
The 53 Hz window of the lowpass filter at the output of lock-in-amplifier allows the entire power line fluctuations (50 Hz) to get
superposed on the sample voltage fluctuations, which reveals itself as peaky structures in the spectrum through aliasing. By increasing $\tau_{lock}$ (thus reducing the width of the pass-band) by a factor of 2, the situation improves only marginally as the roll-off suppressed the 50 Hz noise only by 24 dB, i.e. a factor of only 10. This, however, is not good enough for the power line rejection. In 
addition, this procedure restricts the measurable noise bandwidth from the sample apart from making the whole process rather slow.
The problem was solved by the following digital signal processing scheme. The scheme enables one to suppress the power line fluctuations at 50 Hz by a factor of $\sim 10^{3} - 10^{5}$ keeping the sampling rate even at 32 Hz and other lock-in-amplifier parameters same as described above. 

An example of the result obtained by this process is shown  in figure 4 where we show the data that have been processed by the DSP scheme and that used only the filter of the lock-in amplifier. A comparsion of the two cases shown in figure 4(a) and (b) shows that the background has been suppressed by nearly 3 orders by the DSP technique used here. The observed background noise after the proper signal processing is white with a spectral power of $\approx 10^{-18}$ V$^2$/Hz, which is close to the value of Johnson noise ($4k_{B}TR$) for a 100 ohm resistor at 300K. 
\begin{figure}[h]
\begin{center}
\epsfig{figure=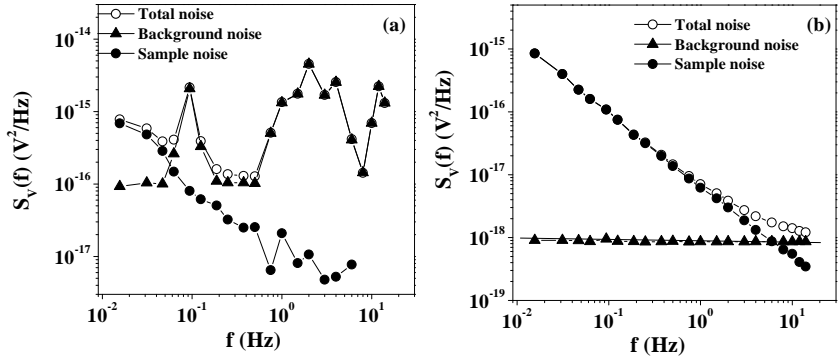,width=12cm,height=8.36cm,clip=}
\vspace{-2cm}
\caption{An example of the power spectra obtained with and without the DSP technique described in the text. (a) The power spectra without the decimation and without DSP techniques (using lock-in filter only) showing large contribution from the power line. (b) Decimation and DSP techniques lead to complete suppression of the power line noise. The observed background is the Johnson noise of the resistor. The data are for a 100 ohm carbon resistor. }
\end{center}
\end{figure}
This example clearly shows that our technique suppresses all extra noise source contribution. The filter in the digital lock-in is not good enough to suppress the background noise to this extent. (Note: The DSP scheme mentioned here takes care of the power line interference as well as the aliasing due to improper digitization rate. The interference due to the power-line is more serious. The spectral power due to the power line at 50Hz, within the shielded enclosure is around 10$^{-12}$ to $10^{-14}$ V$^2$/Hz. The DSP scheme mentioned here suppresses the powerline interference at the frequency window used here ,so that the only contribution to the background is the Johnson noise of the resistor.)

The previous discussion addresses the important problems that the DSP scheme takes care of. Below we describe briefly the  operational part of the scheme. In figure~5 we describe the schematic of the DSP technique used.
\vspace{-2cm}
\begin{figure}[h]
\begin{center}
\epsfig{figure=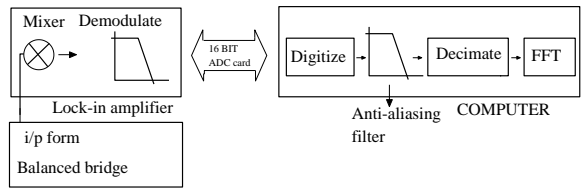,width=8cm,height=8cm,clip=}
\vspace{-2cm}
\caption{Schematic of the digital processing process.}
\end{center}
\end{figure}
 Let $\Delta$ be the intended sampling rate by the DAC. Typically,$\Delta$ = 32, 16, 8,..Hz, usually a power of two, even though it is not absolutely necessary to restrict $\Delta$ only to such
values. The lock-in-amplifier output filter time constant is kept
at the nearest available value of $1/2\pi\Delta$ and a roll-off 24
dB/oct. The data processing scheme goes as follows~\cite{ag1}:

(i) Raw data is digitized at a rate of $\Delta_{d}$ = 16$\Delta$
or 32$\Delta$. The essential idea is that the gain of the filter
at the lock-in-amplifier output comes down by $\sim$ 96 to 120 dB
(i.e. by factor of $10^5-10^6$) at the digitization rate. This
suppresses any significant aliasing effect in the digitized data
that is stored in the computer memory. 

(ii) A digital antialiasing filter with  pass-band cutoff at
$\Delta$ and a roll-off of $\sim$ 100-120 dB/oct was applied on
the digitized data.

(iii) Finally, after applying the anti-aliasing filter on the raw
data, it is decimated by storing every 16$^{th}$ or 32$^{nd}$ data
points, depending on the ratio of original digitization
rate $\Delta_{d}$ and$\Delta$. The new series of data has an
effective sampling rate of  and $\Delta$ and almost devoid of any
power line disturbances due to high roll-off of attained in the
antialiasing filter.

(iv) Steps (ii) and (iii) taken together is the decimation in time
domain. In actual data processing, one usually decimates in
two/three steps in order to optimize the storage requirement and
computation time. In the next section a three step decimation
process is described with typical numbers encountered during the
experiments.

\section{Digital filters: Mutlistage decimation in time domain}

A schematic of the multistage decimation process is shown in figure 6. 
\begin{figure}[h]
\begin{center}
\epsfig{figure=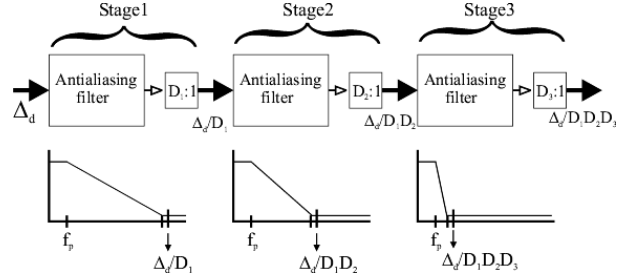,width=8cm,height=3.58cm,clip=}
\vspace{0.5cm}
\caption{Multistage decimation process .}
\end{center}
\end{figure}

For decimation by a total factor of 32, we have used
the set of successive decimation factors of 4,4 and 2, while for a total decimation of 16, factors of 4,2 and 2  were found to be most efficient. The roll-off of the digital filter also varies with the stage of decimation and it is  maximum at the final stage. The final stop-band attenuation is however limited by the attenuation of the lock-in-amplifier output
at the digitization rate $\Delta_d$. Since $\Delta_d \gg \Delta$ there is appreciable attenuation of the extraneous signal at the stop band .

The antialiasing filter design was mainly governed by two requirements: attenuation in the stop-band and transition band-width. All filters employed in decimation procedure are FIR (``finite duration impulse response'')-type whose operation can be
expressed by the following convolution term in time domain~\cite{ag8,numerical,dsp},
\begin{equation}
y[n] = \sum^{N}_{m=0}x[n-m]h[m]
\end{equation}

\noindent where $h[m], m=0,1,..,N$ is the impulse response function of the filter, $N+1$ is the filter order or the number of
taps, $x$ and $y$ are input and output sequences respectively. We adopted the Kaiser window method of  filter designing~\cite{oppenheim}. The Kaiser window is defined as,
\begin{equation}
w[n] = \Bigg\lbrace {\frac{I_0[\beta(1-[(n-\alpha)/\alpha]^2)^{1/2}]}{I_0(\beta)} \qquad  0\leq n\leq M \atop 0  \qquad \qquad \hspace{1.8cm} \mbox{otherwise}}
\end{equation}

\noindent where $\alpha = M/2$ and $I_0$ is the zeroth-order
modified Bessel function of the first kind. By varying $M+1$ and
the shape parameter $\beta$, the window length and shape can be
adjusted to trade sidelobe amplitude for mainlobe width. . As a typical example, for a total decimation by a factor of 16 we used successive stages of 4,4 and 2.The corresponding number of taps (length of antialiasing filters) are 35,25 and 321 to achieve a total attenuation of 100dB at the stop band. The final sampling rate is 4Hz and the maximum frequency limit is limited to 2Hz. The filter response functions are shown in figure~7 along with the stop frequency. In the same graph we show the filter function of the lock-in amplifier. It can be seen that the attenuation attained at the stop band by the digital filters (along with decimation) is much superior to that attended by the lock-in amplifier filter. We refer to figure~4 and note that the 3 order attenuation of the extraneous back ground signal is possible due the filter functions and the decimation process.
\begin{figure}[h]
\begin{center}
\epsfig{figure=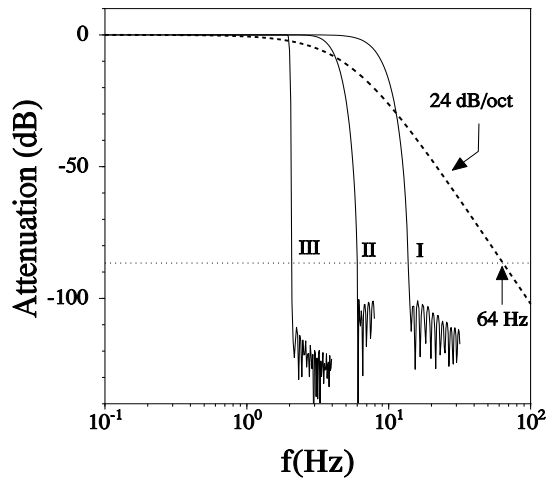,width=8cm,height=7.57cm,clip=}
\caption{The response function of the three stage digital filters used. The lock-in amplifier filter response is shown for comparison as a dotted line.}
\end{center}
\end{figure}

\section{Estimation of the power spectrum: Welch's Periodogram Method}

The digitized and subsequently decimated data was then processed to obtain the power spectrum of conductance fluctuations. The
estimation of the power spectrum used a method developed by Peter D. Welch using fast fourier transform and is known as the ``method of averaged periodogram''~\cite{welch}. The principal advantages of this method are a reduction in the number of computation and required core storage and hence smaller execution time.The essential steps of
the averaged periodogram method are listed below.
Let, $x(j), j = 0,1,...,N-1$, be a sample from a stationary,
second-order stochastic sequence. Let $S_V(f), |f| \leq \Delta/2$,
be the power spectral density of $x(j)$. At First, the data is
divided in segments of length $L$ with the starting point of these
segments $D$ units apart. Then,

\begin{eqnarray}
x_1(j) &=& x(j) \qquad\qquad\qquad\qquad j=0,1,...,L-1\nonumber\\
x_2(j) &=& x(j+D) \qquad\qquad\qquad j=0,1,...,L-1\nonumber\\
.....\nonumber\\
x_K(j) &=& x(j+(K-1)D) \qquad j=0,1,...,L-1,\nonumber
\end{eqnarray}

\noindent $K$ being the total number of segments. Hence, $N = L + (K-1)D$.  A modified periodogram for each segment $x_{k}(j)$ is calculated by the method described below.The schematic of overlapping data segmentation  is shown  in figure 8. 
\begin{figure}[h]
\begin{center}
\epsfig{figure=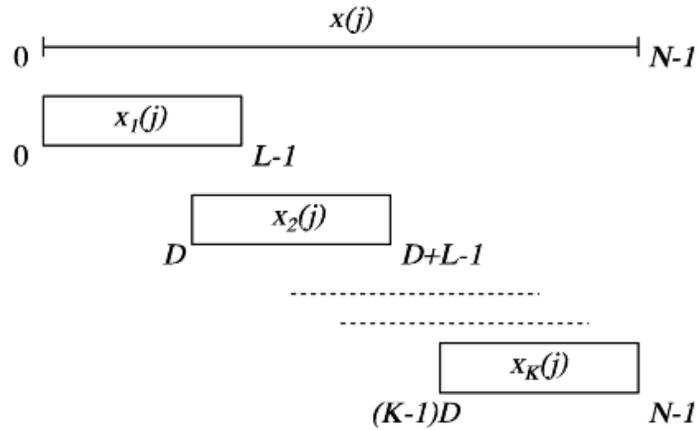,width=10cm,height=6.3cm,clip=}
\vspace{0cm}
\caption{The schematic of the data segmentation used to  construct the periodogram which gives the discrete FFT .}
\end{center}
\end{figure}

A window function $w(j), j=0,1,...,L-1$, is selected and multiplied to $x_{k}(j)$ to form the sequences $x_1(j)w(j),  x_2(j)w(j),...,x_K(j)w(j)$. Discrete Fourier transforms  $A_1(n), A_2(n),...,A_K(n)$ of the K sequences 
${x_1, x_2....x_k}$ and for n=0,1,2 .....L/2 can be expressed as
\begin{equation}
A_k(n) = \frac{1}{L}\sum^{L-1}_{j=0}x_k(j)w(j)e^{-2kijn/L}
\end{equation}
\noindent $i = \sqrt{-1}$. The modified periodograms are obtained as,
\begin{equation}
P_k(f_n) = \frac{L}{U}|A_k(n)|^2 \qquad k = 0,1,...,K
\end{equation}
\noindent where $f_n = n\Delta/L$, for $n = 0,1,...,L/2$ and $U = (1/L)\sum\limits^{L-1}_{j=0} w^2(j)$. U is the normalization
factor . The discrete frequency steps in the power spectrum are in multiples of $\Delta/L$. The maximum frequency is limited by the
effective sampling rate and is $\Delta/2$. While $\Delta$ is the final effective data taking rate after the decimation, the length
of each sequence L can be varied after the data are taken and an optimum value of $\Delta/L$ can be reached that will  also
minimize the speed of computation. Since N is fixed the data analysis has two out of the three parameters L,K and D free.
Generally in the program the input L and D are put by hand during the calculation at the final stage.

The spectral estimate is the average of these periodograms, i.e., 
\begin{equation}
\hat{S}(f_n) = \frac{1}{K}\sum^K_{k=1}P_k(f_n)
\end{equation}
\noindent The largest frequency of spectrum is limited to $\Delta$/2, $\Delta$ being the sampling rate, by Nyquist theorem. In most case of estimation of power spectral density, ``Hanning window function'' which has the following form was used for $w(j)$ as~\cite{ag8,numerical,dsp},
\begin{equation}
w(j) = \frac{1}{2}[1 - cos(\frac{2\pi j}{N})] \qquad j = 0,1,...,L-1
\end{equation}
\noindent The program to evaluate the spectral estimate is mostly taken from standard sources (~\cite{numerical})with a few
modifications. For any power spectral density estimate, checking the normalization is absolutely necessary. The normalized
estimation of the power spectral density evaluated in the method described so far has been cross-checked with the estimation of the
same quantity using commercially available FFT program. All the evaluations gave the same estimate of $S_V(f)$. (Note: While there are a number of other ways the power spectra can be evaluated, this is one of the simplest to implement with a well-defined expression for the normalization of the estimated power spectrum. The variance of the estimator of the power spectrum is also somewhat small in Welch's periodogram method.)

\section{Analysis in the  time domain : Weiner Filter}

So far our discussion has been focused on obtaining the power spectra from the time series (which is the actual stored signal) using various tools of DSP. In certain cases, it is of interest to study the time series of voltage fluctuations directly and obtain information from it. In particular, it is very important for calculating the higher order statistics, as we see in the next sections. In the case when the conductance fluctuations are orders of magnitude larger than the background level, the time domain analysis is usually accurate if we use only the ``in-phase" time series as it is recorded from the lock-in amplifier output. However, if the conductance fluctuations are such that within the measured bandwidth, it is comparable in magnitude with the background, then the ``in-phase" time series is sufficiently corrupted by the additional background signals, and the effect of the background has to be removed before time domain analysis can be done. We describe a method to obtain the uncorrupted time series from the recorded time series of voltage fluctuation. In our method for obtaining the ``uncorrupted" time series, the data from the ``in-phase" as well as the ``quadrature" components recorded by the lock-in amplifier are required. The former contains signal from the conductance fluctuations as well as the background fluctuations, while the latter contains only the background fluctuations. The method of estimating the power spectrum is a powerful tool for obtaining this ``uncorrupted" signal from the corrupted one. We have used Weiner filter~\cite{numerical} to eliminate the background signal. The Weiner filter translates the signals into frequency domain, and then retains only that fraction of the power that corresponds to the ``excess" noise over the background. It then re-transforms the ``cleaned" spectrum into the time domain to produce an ``uncorrupted" time domain signal. The power spectrum of the uncorrupted time domain signal contains only the ``excess" noise, without the background being present.
 Below, a brief description of the adopted procedure is given.

  Let $c(t)$,$n(t)$ and $s(t)$ be the corrupted, background and the uncorrupted time domain signals respectively. Let $C(f)$, $N(f)$ and $S(f)$ be their Fourier transforms, respectively. We define the filter transfer function in the frequency domain,
\begin{equation}
\Phi(f) = \frac{\vert S(f)\vert ^2}{\vert S(f)\vert ^2 + \vert N(f)\vert ^2}
\end{equation}
This is the fraction of power, at each frequency, of the uncorrupted signal in the total signal. Since $\vert C(f)\vert ^2 = \vert S(f)\vert ^2 + \vert N(f)\vert ^2$, we can rewrite this as
\begin{equation}
\Phi(f) = \frac{\vert C(f)\vert ^2 - \vert N(f)\vert ^2}{\vert C(f)\vert ^2}
\end{equation}
To obtain the uncorrupted signal, the corrupted ``in-phase" signal is first Fourier transformed. It is then multiplied with this filter transform function and inverse Fourier transformed to give the ``uncorrupted" signal.An important assumption in this estimation is that the background and the uncorrupted signals are uncorrelated.
Figure~9 shows a flow-chart describing this procedure. 
\vspace{-0.5cm}
\begin{figure}[h]
\begin{center}
\epsfig{figure=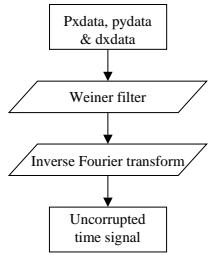,width=8.12cm,height=5.66cm,clip=}
\vspace{-1cm}
\caption{Flow chart showing the procedure for obtaining the ``uncorrupted" signal. {\tt pxdata} and {\tt pydata} corresponds to the power spectra of the in-phase and quadrature signals. {\tt dxdata} is the (decimated) corrupted time domain signal.}
\end{center}
\end{figure}
In figure~10, we show how a typical time series looks like before and after Weiner filtering. The highest frequency components get suppressed sharply since they are dominated by the background noise. Besides its usefulness in studying the time domain features, the estimation of the uncorrupted signal is crucial for the other statistical and spectral analysis that were extensively used in our investigations.
\begin{figure}[h]
\begin{center}
\epsfig{figure=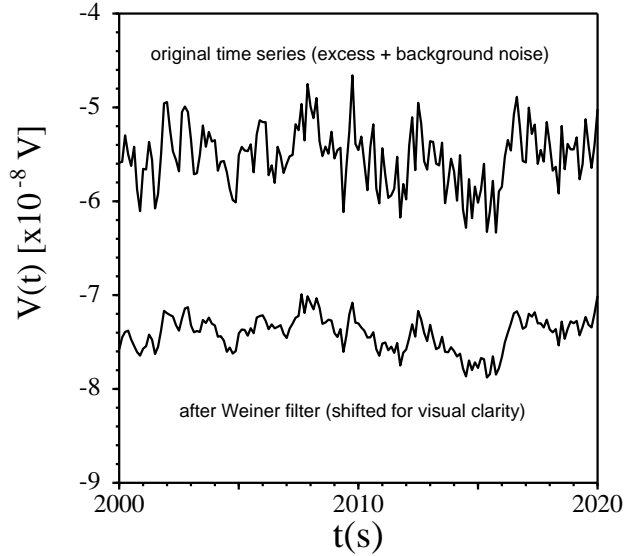,width=9cm,height=9cm,clip=}
\vspace{0cm}
\caption{A typical time series of voltage fluctuations before and after Weiner filtering. Note that the low frequency fluctuations, where the spectral power of the ``1/f" noise dominates, are almost unchanged , whereas the high frequency fluctuations where the spectral power of ``1/f" noise is negligible and the total fluctuations are dominated by the background noise, are sharply attenuated. The filtered time series has been slightly shifted for better viewing.}
\end{center}
\end{figure}

\section{Calculation of Probability Density function}

It is a common practice to characterize noise by its power spectral density, $S_V(f)$. In the context of Gaussian fluctuation this will suffice. In many physical situation, however, the fluctuation need not be Gaussian. In some complicated systems, the fluctuators may not be completely independent of each other, giving rise to correlations. This kind of behavior is expected from systems with extremely large relaxation time or with sequential kinetics with hierarchy as envisaged in frustrated systems like spin glass of structural glass freezing.~\cite{weissman2} Under these circumstances,limiting the analysis of conductance fluctuations to its spectral power density only does not completely characterize the fluctuating system. When the fluctuations become non-Gaussian, and higher order statistics are needed  to describe the fluctuations. Mathematically, Gaussianity implies that all multi-point correlation functions $\langle V(t)V(t+t_1)...V(t+t_n)\rangle$ can be expressed by decomposing them into the sum of all possible products of pairwise correlation functions~\cite{bendat,nelkin}. Another way to check the Gaussian nature of the fluctuation or deviation from it is to determine the PDF (the fluctuation histogram) from the time series $\delta v(t)$. The PDF of voltage fluctuations   had been investigated even more than 3 decades back to study the statistical nature of the $1/f$ noise in resistors and semiconductor devices ~\cite{brophy}.

We have followed the following procedure to obtain the PDF from the observed time series~\cite{ag8,sk4}. This is a markedly improved procedure than past practices and allows determination of PDF when the relative variance of the fluctuation is less than even 1 part in 10$^4$. In the first step, the power spectrum is calculated to find out the comparative levels of excess noise present with respect to the background noise. This can be done by studying the time series obtained from the in-phase and the quadrature output of the lock-in amplifier. This is important in order to assess to what extent  that the background effect can affect the signals.  Next we find the  approximate frequency ($f^*$) at which the excess noise from the sample (which is often of $1/f$ nature) and the background noise become comparable.For example it can be seen that in ~figure 4 $f^{*}\approx 6Hz$ for the carbon resistor used. If the sample has very low noise level (an example will be given in figure~13) then $f^*$ can be indeed be very low like $f^{*}\approx 0.1Hz$.$f^*$ fixes the bandwidth.The real time signal is decimated down so that the bandwidth is restricted below $f^*$ . In this way, we ascertain the in-phase component is always above the background level. For $f > f^*$  the signal is sufficiently corrupted that the filtering cannot remove the background. The in-phase component time series (decimated to $f < f^*$) is Weiner filtered in the method described in the previous section, and we get the uncorrupted time domain signal without significant contribution from the background level.
\vspace{-0.5cm}
\begin{figure}[h]
\begin{center}
\epsfig{figure=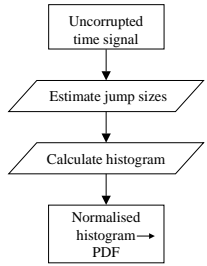,width=8cm,height=5.6cm,clip=}
\vspace{-1cm}
\caption{Flow chart showing the procedure for obtaining the PDF.}
\end{center}
\end{figure}
From the uncorrupted time series the PDF is then calculated in two steps.In the first step, the minimum voltage jump size ($\delta v_m$)is estimated from the lowest level of voltage fluctuations in the power spectrum . As stated this is mostly the $4k_{B}TR$ background noise in many cases. Formally $\delta v_{m}^2\approx \int_{0}^{f^*}S_{background}(f)df$ . The jump size of each fluctuation from the uncorrupted time series is calculated and stored in a separate file, ignoring all jumps with $\vert \delta v\vert \leq \vert \delta v_{m}\vert $, the minimum assigned voltage jump size. Effectively, $\delta v_m$  sets the ``zero" of the PDF.

Next, voltage ``bins" are created, starting from the minimum voltage jump size upto the maximum jump value obtained from the time series. The voltage jump sizes stored in a separate file are then assigned into corresponding voltage bins. The ensemble of these bins with the number of ``jumps" in each voltage bin makes  the histogram for the fluctuation. The histogram is normalized so that the total area under the distribution is unity, and this normalized histogram is the PDF. This procedure has been shown through a flowchart in figure~11. 

\section{Calculation of ``Second Spectrum"}
Deviation of the PDF (as calculated in the section before) from a Gaussian is a way to represent non-Gaussian behavior of a fluctuating system. A second way to do this is to measure the variance of the noise power determined from a series of measurements within a given bandwidth and to compare it with that expected from Gaussian noise~\cite{weissman2,restle}. This can be further extended to calculate the variance as well as the covariance and generate the covariance matrix. For Gaussian noise, the covariance elements of this matrix have expectation values of zero. An equivalent way of expressing this information is to take the power spectra and cross spectra of the time series of noise power measurements in one or more bands. This power spectrum of the fluctuations in noise power within a band of the original spectrum is called the Second Spectrum. The second spectrum is the FFT of the four point correlation function expressed as~\cite{solin1}  $S^{(2)}(f)$ = $\int_{0}^{\infty} \langle \rm{V}^2(t)\rm{V}^2(t+\tau)\rangle_t cos(2\pi f\tau)d\tau$. To make an estimate the second spectrum , we start with the uncorrupted signal. This is then digitally band-passed through a small frequency window, within which the fluctuation of the spectrum is to be tested. The data sequence is then squared, point-by-point and stored as another file. The power spectrum of this ``squared" time series gives the second spectrum~\cite{sk4}. The figure~12 shows the steps for estimating in  the second spectrum.In the section below we give examples of second spectra taken by our program.
\vspace{0cm}
\begin{figure}[h]
\begin{center}
\epsfig{figure=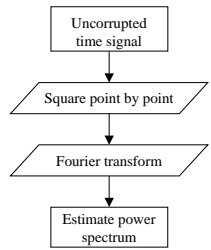,width=8.1cm,height=5.6cm,clip=}
\vspace{-1cm}
\caption{Flow chart showing the procedure for obtaining the second spectrum.}
\end{center}
\end{figure}

\section{Examples of noise measurements}

This apparatus has been used in a number of systems for noise measurements.~\cite{ag1,ag2,ag3,ag4,ag5,ag6,sk1,akr2,ag7,sk2,ab1,sk3} In the following we quote some examples from them and refer to the original paper for details. An example of a typical $1/f$ noise measured in the apparatus is shown in figure~13(a) for a Ag film (1.5mm x 12.5$\mu$m x 100nm) grown on glass substrates. 
\begin{figure}[h]
\begin{center}
\epsfig{figure=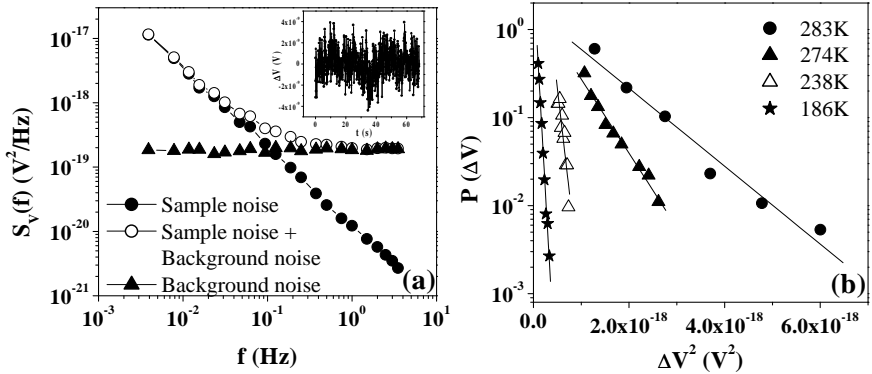,width=12cm,height=8.3cm,clip=}
\vspace{-1.5cm}
\caption{(a)Example of spectral power and time series  (inset) in a low noise Ag film. (b) The PDF of the voltage fluctuation(obtained from the time series after Wiener filtering as explained in the text). The PDF in this case is a Gaussian. ( Note :  log($\Delta V$) is plotted against  $\Delta V^2$ to display the Gaussian nature of the PDF.) The width increases as the temperature is raised.}
\end{center}
\end{figure}
The inset shows the time series of the fluctuations. The spectral noise measured has $1/f$ dependence and at $f\approx 1 Hz$ the spectral power (after background subtraction)is as low as $10^{-20}$ V$^2$/Hz. This is lower than the background noise of $10^{-19}$ V$^2$/Hz which is close to the Johnson limit. In  figure~13(b) we show the PDF of the time series obtained after Wiener filtering explained above.The Gaussian nature of the fluctuation can be seen from the PDF. The width of the PDF increases on heating following an activated dependence on temperature with an activation energy of $\approx 0.3eV$.

Figure 14 shows typical spectra obtained in P doped silicon at low temperature (4.2K). 14(a) Shows the spectrum taken in a sample of low resistance ($\sim 10\Omega$), where a transformer preamplifier was used before feeding the signal into the lock-in amplifier. The resulting background noise level is very low. In 14(b), we see the spectrum taken in a sample of higher sample resistance ($\sim 100\Omega$) and the signal was fed directly into the lock-in amplifier. As a result, the background noise level was much higher. In spite of this high background, we can see remarkably clear ``1/f" behavior of the spectrum upto two order of magnitude below the background noise level. This level of clarity was possible only due to the flexibility obtained in signal recovery using DSP methods.
\begin{figure}[h]
\begin{center}
\epsfig{figure=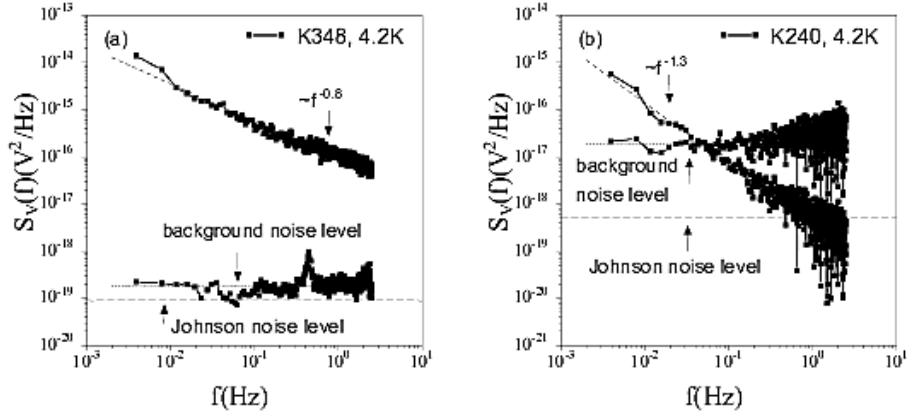,width=12cm,height=5.74cm,clip=}
\caption{(Spectra taken in two representative samples of P doped Si at 4.2K. (a) A low resistance sample allowing the usage of a transformer preamplifier, lowering the noise background. (b) A high resistance sample, signal from which was directly fed into the lock-in, giving rise to higher background level. Note the clarity with which the ``1/f" part is visible even two orders of magnitude below the background level. The data are from references [18] and [38].}
\end{center}
\end{figure}
As mentioned before, our setup allows us to measure noise over large ranges of temperature. An nice example is noise measured in P doped Si from 5K to 500K. Figure 15 shows the scaled noise parameter $\gamma_H$ as a function of T. 
\begin{figure}[h]
\begin{center}
\epsfig{figure=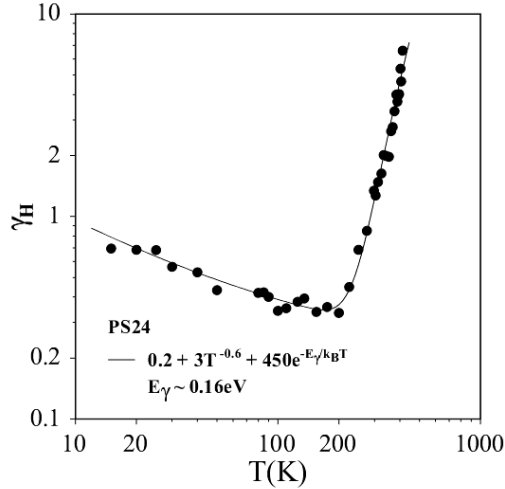,width=7cm,height=7cm,clip=}
\vspace{0cm}
\caption{T dependence of the scaled noise parameter $\gamma_H$ in P doped Si, clearly showing various (power law, activated) behavior. }
\end{center}
\end{figure}
The scaled noise parameter $\gamma_H$ ( often called the Hooge parameter)  is defined as 
\begin{equation}
\gamma_H = \frac{f^\alpha S_V(f)N}{V^2}
\end{equation}
where V is the mean bias across the sample containing N charge carriers and $\alpha \simeq 1$. The noise at low temperatures ($T < 100$ K) rises with decrease in T following an inverse power-law, as determined by the mechanism of Universal Conductance Fluctuations (UCF)~\cite{ag4}. Close to the critical regime of the metal-insulator transition in doped Si the noise diverges as the fluctuation becomes non-Gaussian as revealed by higher cumulants like the second spectra or probability density function obtained from the real time series~\cite{sk3}. At higher temperatures (T$>$100K), fluctuation arises from different mechanisms and has an activated dependence on temperature. 

Perhaps one of the biggest successes of our setup was the ability to systematically study not only the fluctuation spectra and statistics, but also clean measurements of the second spectrum in doped silicon which has rather low noise levels ($<\Delta R>$/R $<$ $10^{-4}$)~\cite{sk3}. In figure 16 we show the second spectrum measured in a doped Si sample at various temperatures. The systematic growth in the slope of the spectrum shows clear indication of growth of correlations within the fluctuators.~\cite{sk3}
\begin{figure}[h]
\begin{center}
\epsfig{figure=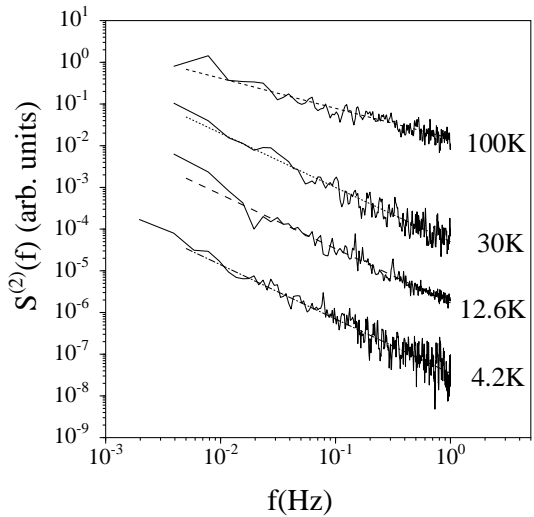,width=8cm,height=8cm,clip=}
\vspace{0cm}
\caption{Second spectra of fluctuations in P doped Si, measured at various temperatures. The growth in the slope of the second spectra indicates correlations (see text). }
\end{center}
\end{figure}

A nice example of measurements with a.c with  superimposed d.c is the non-linear conduction and the accompanied noise in charge ordered systems like the rare-earth manganites~\cite{ab1}. This is  single crystal of the charge order manganite Pr$_{0.63}$Ca$_{0.37}$MnO$_3$. The charge ordering temperature is 260K. But below the charge ordering temperature ($\approx 220K$), the charge ordered state can be destabilized by an applied d.c bias which leads to non-linear conduction.  
\begin{figure}[h]
\begin{center}
\epsfig{figure=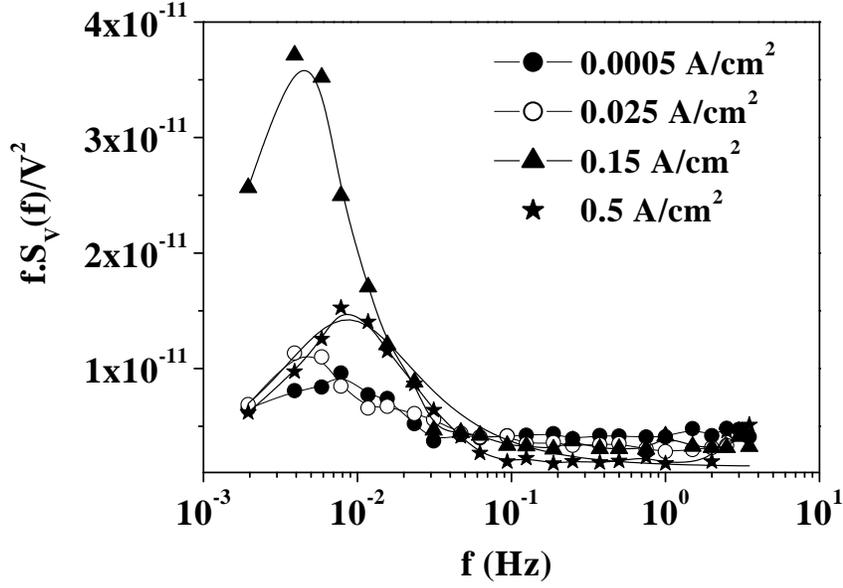,width=12cm,height=8.36cm,clip=}
\vspace{0cm}
\caption{Spectral power of a charge ordering manganite~\cite{ab1}. The spectral noise has a narrow band low frequency noise which increases when an applied d.c bias melts the charge ordering. The noise data are taken with a a.c bias with a superimposed d.c.}
\end{center}
\end{figure}
In figure 17 we show that at the threshold of non-linear conduction the noise contains random telegraphic noise (RTN) which gives distinct Lorentzians superimposed on a broad background of $1/f$ noise.~\cite{ab1} In this case the noise was measured by an a.c in presence of a superimposed d.c bias.

This work has been financially supported by the CSIR, Govt. of India, through an SRF (SK), an SPMF (AB) and a sponsored scheme (AKR).


\begin{references}
\bibitem{vdz} A. Van Der Ziel , Adv. in electronics and electron physics {\bf 24}, 225 (1979).
\bibitem{DH} P.Dutta and P.M. Horn, Rev. Mod. Phys. {\bf 53}, 497 (1981).
\bibitem{weissman1} M. B. Weissman, Rev. Mod. Phys. {\bf 60}, 537 (1988).
\bibitem{weissman2} M. B. Weissman, Rev. Mod. Phys. {\bf 65}, 829 (1993).
\bibitem{bkj} B.K Jones , Adv, in electronics and electron Physics {\bf 87 }, 201 (1994).
\bibitem{akr1} A.K. Raychaudhuri , Current Opinion in Solid State and Materials Science {\bf 60}, 67 (2002).
\bibitem{ag1} A.Ghosh and A.K.Raychaudhuri,R Sreekala , M Rajeswari  and T Venkatesan, J. Phys D: Appl. Phys {\bf 30L},75(1997)
\bibitem{ag2} A.Ghosh and A.K.Raychaudhuri, J.Phys.:Condens. Matter {\bf 11}, L457 (1999).
\bibitem{ag3} A.Ghosh and A.K.Raychaudhuri, Phys. Rev B {\bf 58}, R 14665 (1998).
\bibitem{ag4} A.Ghosh and A.K.Raychaudhuri, Phys. Rev. Letts. {\bf 84},4681 (2000).
\bibitem{ag5} A. Ghosh and A.K. Raychaudhuri, Proc. of the $16^{th}$ International conference on Noise in Physical Systems and 1/f Fluctuaions (World Scientific) 2001, page 107.
\bibitem{ag6} A.Ghosh and A.K.Raychaudhuri, .Phys. Rev  B, {\bf 64},104304 (2001)
\bibitem{sk1} S.Kar and A.K. Raychaudhui, J. Phys. D: Appl. Phys. {\bf 34} 3197 (2001).
\bibitem{akr2} A.K. Raychaudhuri {\it et al.}, Pramana - Journal of Physics {\bf 58}, 343 (2002).
\bibitem{ag7} A.Ghosh and A.K.Raychaudhuri,Phys. Rev. B, {\bf 65}, 033310 (2002) .
\bibitem{sk2} S.Kar and A.K. Raychaudhuri, Appl. Phys.Letts {\bf 81} 5165 (2002).
\bibitem{ab1} A.Bid,A.Guha and A.K. Raychaudhuri, Phys.Rev B, {\bf 67 } 174415(2003).
\bibitem{sk3} S.Kar, A.Ghosh and A.K.Raychaudhuri , Phys. Rev. Letts. {\bf 91},216603(2003)
\bibitem{scoff1} J. H. Scoffield, Rev. Sci. Instr. {\bf 58}, 985 (1987).
\bibitem{verbruggen} A.H. Verbruggen, H.Stoll, K. Heeck and R.H. Koch, Appl. Phys. A {\bf 48}, 233(1989).
\bibitem{reif} Fedrick Reif ``Fundamentals of Statistical and Thermal Physics" McGraw-Hill, New York (1965).
\bibitem{ag8} A.Ghosh  ``Low frequency Conductance Fluctuation near metal-insulator transition " Ph.D thesis (1999) , Indian Institute of Science, Bangalore , India .
\bibitem{Bourns}BOURNS, 1200 Columbia Avenue, Riverside, CA 92507, USA, www.bourns.com
\bibitem{triad}Magnetek Inc.10900 Wilshire Blvd. Suite 850, Los Angeles, CA 90024,USA,www.magnetek.com
\bibitem{srs}Stanford Research Systems, Inc. 1290-D Reamwood Ave, Sunnyvale, CA 94089,USA,www.srsys.com
\bibitem{cooner}Cooner Wire,9265 Owensmouth, Chatsworth,CA 91311,USA,www.coonerwire.com.(Stock number CZ1103-1F)
\bibitem{national instruments}National Instruments Corporation, 11500 N Mopac Expway, Austin , TX 78759,USA,www.ni.com
\bibitem{plc}ADVANTECH, No 1, Alley 20, Lane 26, Rueiguang Road, Neihu District, Taipei 114 , Taiwan, www.advantech.com.tw
\bibitem{numerical}W.H.Press, S.A.Teukolsky, W.T.Vetterling and B.P. Flannery ``Numerical Receipes", Cambridge University Press (1993)
\bibitem{dsp}R.W.Hamming, ``Digital Filters", Englewood Cliffs, N.J, Prentice Hall (1983), D.J.DeFatta, J.G.Lucas and W.S.Hodgkins ``Digital Signal Processing" , John Wiley and Sons , NY (1988)
\bibitem{oppenheim}A.V.Oppenheim and R.W. Schafer, ``Discrete Signal Processing", Englewood Cliffs, N.J, Prentice Hall (1989).
\bibitem{welch}P.D. Welch in ``Modern Spectral Analysis", p17 Ed. D.G.Childers, IEEE press, John Wiley and Sons , NY (1978)
\bibitem{bendat}J.S.Bendat and A.G.Piersol, ``Random Data:Analysis and Measurement Procedures", Wiley , NY (1986).
\bibitem{nelkin}M.Nelkin and A.M.S. Tremblay, J. of Stat. Phys.{\bf 25}, 253 (1981)
\bibitem{brophy}J.J.Brophy, Phys. Rev. {\bf 166}, 827 (1968) and J.Appl.Phys.{\bf 40}, 3551 (1969).
\bibitem{restle}P.J.Restle, M.B.Weissman and R.D.Black,J.Appl.Phys.{\bf 54}, 5844 (1983).
\bibitem{solin1} G.T.Seidler and S.A Solin, Phys. Rev. B, {\bf 53}, 9753 (1996).
\bibitem{sk4}S.Kar  ``Spectral analysis of Conductance Fluctuations in doped Silicon near the Mott-Anderson transition" Ph.D thesis (2003) , Indian Institute of Science, Bangalore , India .

\end{references}
\end{document}